\documentclass[twocolumn,aps,prc,showpacs,superscriptaddress,nofootinbib,floatfix]{revtex4}
\usepackage{graphicx}
\usepackage{dcolumn}
\usepackage{bm}
\usepackage{longtable}
\begin{document}

\title{Clustering of Color Sources and the Shear Viscosity of the QGP in Heavy Ion collisions  at RHIC and LHC Energies}
\affiliation{CENTRA, Instituto Superior Tecnico, 1049-001 Lisboa, Portugal}
\affiliation{Department of Physics, Purdue University,West Lafayette, IN-47907, USA}
\affiliation{Departamento de Fisica de Particulas, Universidale de Santiago de Compostela and Instituto Galego de Fisica de Atlas Enerxias(IGFAE), 15782 Santiago, de Compostela, Spain}
\author{J. Dias de Deus}\affiliation{CENTRA, Instituto Superior Tecnico, 1049-001 Lisboa, Portugal}
\author{A. S. Hirsch}\affiliation{Department of Physics, Purdue University,West Lafayette, IN-47907, USA}
\author{C. Pajares}\affiliation{Departamento de Fisica de Particulas, Universidale de Santiago de Compostela and Instituto Galego de Fisica de Atlas Enerxias(IGFAE), 15782 Santiago, de Compostela, Spain}
\author{R. P. Scharenberg}\affiliation{Department of Physics, Purdue University,West Lafayette, IN-47907, USA}
\author{B. K. Srivastava}\affiliation{Department of Physics, Purdue University,West Lafayette, IN-47907, USA}
\date{\today}

\begin{abstract}
 We present our results on the shear viscosity to entropy ratio ($\eta/s$) in the framework of the clustering of the color sources  of the matter produced at RHIC and LHC energies. The onset of de-confinement transition is identified by the spanning percolating cluster in  2D percolation. 
The relativistic kinetic theory relation for $\eta/s$ is evaluated using the initial temperature($\it T$) and the mean free path ($\lambda_{mfp}$). The analytic expression for $\eta/s$ covers a wide temperature range. At $\it T$ $\sim$ 150 MeV below the hadron to QGP transition temperature of $\sim~$ 168 MeV, with increasing temperatures the $\eta/s$ value drop sharply  and reaches a broad minimum  $\eta/s$ $\sim$ 0.20 at $\it T$ $\sim$ 175-185 MeV. Above this temperature $\eta/s$ grows slowly. The measured values of $\eta/s$ are 0.204$\pm$0.020 and 0.262$\pm$0.026 at the initial temperature of 193.6$\pm$3 MeV from central Au+Au collisions at $\sqrt{s_{NN}}$= 200 GeV (RHIC) and 262.2 $\pm$13 MeV in central Pb+Pb collisions at $\sqrt{s_{NN}}$ = 2.76 TeV (LHC). These $\eta/s$ values are 2.5 and 3.3 times the AdS/CFT conjectured lower bound $1/4\pi$ but are consistent with theoretical $\eta/s$ estimates for a strongly coupled QGP. 
\end{abstract}

\maketitle    
\section{Introduction}
The observation of the large elliptic flow at RHIC in non-central heavy ion collisions suggest that the matter created is a nearly perfect fluid with a very low shear viscosity \cite{brahms,phobos,star,phenix}. Recently, attention has been focused on the shear viscosity to entropy density ratio $\eta/s$  as a measure of the fluidity \cite{teaney,teaney1,lacey,rom}. The observed temperature averaged $\eta/s$, based on viscous hydrodynamics analyses of RHIC data, are suggestive of a strongly coupled plasma \cite{gul1,larry}. The effect of the bulk viscosity is expected to be negligible. 
It has been conjectured, based on infinitely coupled super-symmetric Yang-Mills (SYM) gauge theory using the correspondence between Anti de-Sitter(AdS) space and conformal field theory (CFT), that the lower bound  for $\eta/s$ is $
1/4\pi$ and is the universal minimal viscosity to entropy ratio even for QCD \cite{kss}. However, there are theories in which this lower bound can be violated \cite{buchel}. 
In this work, we use the color string percolation model (CSPM) \cite{pajares1,pajares2} to obtain $\eta/s$ as a function of the temperature above and below the hadron to QGP transition. The measured  $\eta/s$ values are for Au+Au collisions at $\sqrt{s_{NN}}$ = 200 GeV at RHIC and for Pb+Pb collisions at $\sqrt{s_{NN}}$ = 2.76 TeV at LHC. 
\section{Clustering of Color Sources}
Multiparticle production is currently described in terms of color strings stretched between the projectile and the target, which decay into new strings and subsequently hadronize to produce observed hadrons. Color strings may be viewed as small areas in the transverse plane filled with color field created by colliding partons. With growing energy and size of the colliding system, the number of strings grows, and they start to overlap, forming clusters, in the transverse plane very much similar to disks in two dimensional percolation theory. At a certain critical density a macroscopic cluster appears that marks the percolation phase transition. This is the Color String Percolation Model (CSPM) \cite{pajares1,pajares2}. The interaction between strings occurs when they overlap and the general result, due to the SU(3) random summation of charges, is a reduction in multiplicity and an increase in the string tension hence increase in the average transverse momentum squared, $\langle p_{t}^{2} \rangle$. 
 We assume that a cluster of $\it n$ strings that occupies an area of $S_{n}$ behaves as a single color source with a higher color field $\vec{Q_{n}}$ corresponding to the vectorial sum of the color charges of each individual string $\vec{Q_{1}}$. The resulting color field covers the area of the cluster. As $\vec{Q_{n}} = \sum_{1}^{n}\vec{Q_{1}}$, and the individual string colors may be oriented in an arbitrary manner respective to each other , the average $\vec{Q_{1i}}\vec{Q_{1j}}$ is zero, and $\vec{Q_{n}^2} = n \vec{Q_{1}^2} $.
 

Knowing the color charge $\vec{Q_{n}}$ one can obtain the multiplicity $\mu$ and the mean transverse momentum squared $\langle p_{t}^{2} \rangle$ of the particles produced by a cluster of $\it n $ strings \cite{pajares2}
\begin{equation}
\mu_{n} = \sqrt {\frac {n S_{n}}{S_{1}}}\mu_{0};\hspace{5mm}
\langle p_{t}^{2} \rangle = \sqrt {\frac {n S_{1}}{S_{n}}} {\langle p_{t}^{2} \rangle_{1}}
\end{equation} 
where $\mu_{0}$ and $\langle p_{t}^{2}\rangle_{1}$ are the mean multiplicity and $\langle p_{t}^{2} \rangle$ of particles produced from a single string with a transverse area $S_{1} = \pi r_{0}^2$. For strings  just touching each other $S_{n} = n S_{1}$, and $\mu_{n} = n \mu_{0}$, $\langle p_{t}^{2}\rangle_{n}= \langle p_{t}^{2}\rangle_{1}$. When strings fully overlap, $S_{n} = S_{1}$  and therefore 
 $\mu_{n} = \sqrt{n} \mu_{0}$ and $\langle p_{t}^{2}\rangle_{n}= \sqrt{n} \langle p_{t}^{2}\rangle_{1}$, so that the multiplicity is maximally suppressed and the $\langle p_{t}^{2}\rangle_{n}$ is maximally enhanced. This implies a simple relation between the multiplicity and transverse momentum $\mu_{n}\langle p_{t}^{2}\rangle_{n}=n\mu_{0}\langle p_{t}^{2}\rangle_{1}$, which means conservation of the total transverse momentum produced. 

 In the thermodynamic limit, one obtains an analytic expression \cite{pajares1,pajares2}
\begin{equation}
\langle \frac {n S_{1}}{S_{n}} \rangle = \frac {\xi}{1-e^{-\xi}}\equiv \frac {1}{F(\xi)^2}
\end{equation}
where $F\xi)$ is the color suppression factor. With $F(\xi)\rightarrow 1$ as $\xi \rightarrow 0$ and $F(\xi)\rightarrow 0$ as $\xi \rightarrow \infty $, where 
$\xi = \frac {N_{s} S_{1}}{S_{N}}$ is the percolation density parameter. 
Eq.(1) can be written as $\mu_{n}=n F(\xi)\mu_{0}$ and 
$\langle p_{t}^{2}\rangle_{n} ={\langle p_{t}^{2} \rangle_{1}}/F(\xi)$.  
The critical cluster which spans $S_{N}$, appears for $\xi_{c} \ge$ 1.2 \cite{satz1}. 
It is worth noting that CSPM is a saturation model similar to the Color Glass Condensate (CGC),
 where $ {\langle p_{t}^{2} \rangle_{1}}/F(\xi)$ plays the same role as the saturation momentum scale $Q_{s}^{2}$ in the CGC model \cite{cgc,perx}. 

\section{Experimental Determination of the Color Suppression Factor  $F(\xi)$}
The suppression factor is determined by comparing the $\it pp$ and  A+A transverse momentum spectra. 
To evaluate the initial value of $\xi$ from data for Au+Au collisions, a parameterization of $\it pp$ events at 200 GeV  is used to compute the $p_{t}$ distribution \cite{nucleo,levente,eos}
\begin{equation}
dN_{c}/dp_{t}^{2} = a/(p_{0}+p_{t})^{\alpha}
\end{equation}
where a is the normalization factor.  $p_{0}$ and $\alpha$ are parameters used to fit the data. This parameterization  also can be used for nucleus-nucleus collisions to take into account the interactions of the strings \cite{pajares2}
\begin{equation}
dN_{c}/dp_{t}^{2} = \frac {a'}{{(p_{0}{\sqrt {F(\xi_{pp})/F(\xi)}}+p_{t})}^{\alpha}}
\end{equation}
The color suppression factor $F(\xi)$ is related to the percolation density parameter $\xi$.
\begin{equation}
F(\xi) = \sqrt \frac {1-e^{-\xi}}{\xi}
\end{equation}

In pp collisions $F(\xi) \sim$ 1 at these energies due to the low overlap probability.

 In this way the STAR analysis of charged hadrons obtained the preliminary results for the percolation density parameter, $\xi$ at RHIC for several collisions systems as a function of centrality \cite{nucleo}. 
Figure 1 shows a plot of $F(\xi)$ as a function of charged particle multiplicity per unit transverse area $\frac {dN_{c}}{d\eta}/S_{N}$ for Au+Au collisions at 200 GeV for various centralities for the STAR data \cite{levente,eos}. The error on  $F(\xi)$ is $\sim 3\%$. 
$F(\xi)$ decreases in going from peripheral to central collisions. The $\xi$ value is obtained using Eq. (5), which increases with the increase in centrality. The fit to the Au+Au points has the functional form 

\begin{equation}
F(\xi) = exp [-0.165 -0.094 \frac {dN_{c}}{d\eta}/S_{N}]
\end{equation}

\begin{figure}[thbp]
\centering        
\vspace*{-0.2cm}
\includegraphics[width=0.50\textwidth,height=3.0in]{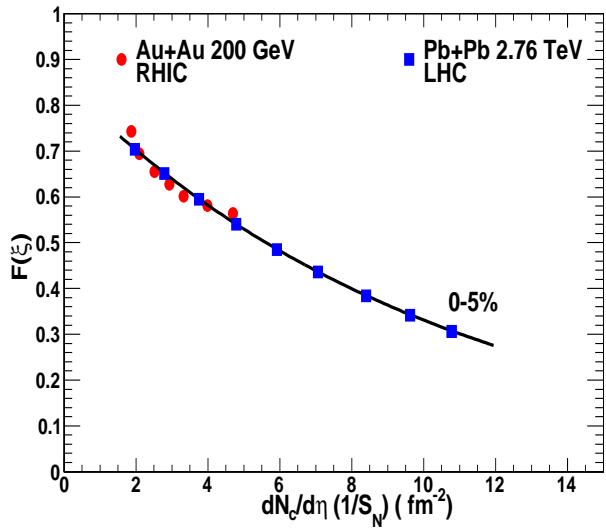}
\vspace*{-0.1cm}
\caption{ Color suppression factor $F(\xi)$ as a function of $\frac {dN_{c}}{d\eta}/S_{N}(fm^{-2})$. 
The solid red circles are for Au+Au collisions at 200 GeV(STAR data) \cite{nucleo}. The error is smaller than the size of the symbol. The line is fit to the STAR data. The solid blue squares are for  Pb+Pb at 2.76 TeV.} 
\end{figure}  

  The STAR results for Au+Au collisions at $\sqrt{s_{NN}}$ = 200 GeV can be used to estimate F$(\xi)$ values for Pb+Pb collisions at different centralities using the fit function given by Eq.(6) for Au+Au.  Recently, the ALICE experiment at LHC published the charged-particle multiplicity density data as a function of centrality in Pb+Pb collisions at $\sqrt{s_{NN}}$ = 2.76 TeV \cite{alice1}. The ALICE data points are shown in Fig.1. For central 0-5$\%$ in Pb+Pb collisions  $\xi$ = 10.56 as compared to $\xi$ = 2.88 for central Au+Au collisions at 200 GeV. 
 For Au+Au central collisions we have found that the Bjorken energy density $\varepsilon$ in the collision is proportional to $\xi$.  To evaluate $\varepsilon$ the charged pion multiplicity at mid rapidity and the Schwinger QED2 production time were used \cite{eos,bjorken}. Figure 2 shows a plot of energy density as a function of $\xi$. $\varepsilon = 0.788 \xi$ for the range 1.2 $ < \xi < 2.88 $. The extrapolated value of $\varepsilon$ for central Pb+Pb collision at 2.76 TeV is  8.32 $GeV/fm^3$ as  shown in Fig.2. 


\begin{figure}[thbp]
\centering        
\vspace*{-0.2cm}
\resizebox{0.55\textwidth}{!}{
  \includegraphics{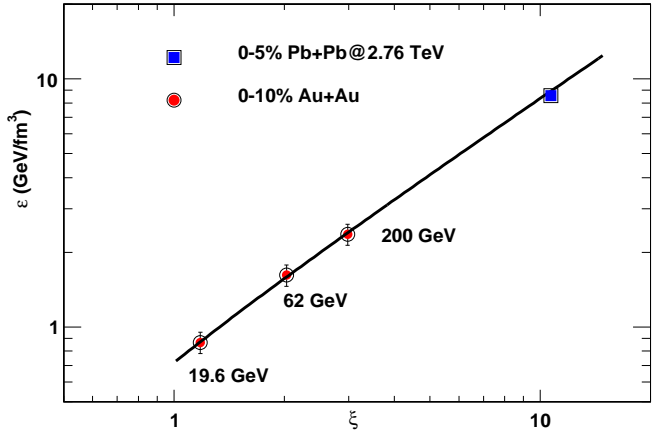}
}
\vspace*{-0.5cm}
\caption{Energy density $\epsilon$ as a function of the percolation density parameter $\xi$. The value for LHC energy is shown as blue square.} 
\end{figure}

\section{Determination of the Temperature}
The connection between the measured $\xi$ and the temperature $T(\xi)$ involves the Schwinger mechanism (SM) for particle production. 
The Schwinger distribution for massless particles is expressed in terms of $p_{t}^{2}$ \cite{swinger,wong}
\begin{equation}
dn/d{p_{t}^{2}} \sim e^{-\pi p_{t}^{2}/x^{2}}
\end{equation}
where the average value of the string tension is  $\langle x^{2} \rangle$. The tension of the macroscopic cluster fluctuates around its mean value because the chromo-electric field is not constant.
The origin of the string fluctuation is related to the stochastic picture of 
the QCD vacuum. Since the average value of the color field strength must 
vanish, it can not be constant but changes randomly from point to point \cite{bialas}. Such fluctuations lead to a Gaussian distribution of the string tension for the cluster, which transforms SM into the thermal distribution \cite{bialas}
\begin{equation}
dn/d{p_{t}^{2}} \sim e^{(-p_{t} \sqrt {\frac {2\pi}{\langle x^{2} \rangle}} )}
\end{equation}
with $\langle x^{2} \rangle$ = $\pi \langle p_{t}^{2} \rangle_{1}/F(\xi)$. 

The temperature is expressed as \cite{pajares3}  
\begin{equation}
T(\xi) =  {\sqrt {\frac {\langle p_{t}^{2}\rangle_{1}}{ 2 F(\xi)}}}
\end{equation} 
 Recently, it has been suggested that fast thermalization in heavy ion collisions can occur through the existence of an event horizon caused by a rapid de-acceleration of the colliding nuclei \cite{khar2}. The thermalization in this case is due to the Hawking-Unruh effect \cite{hawk,unru}. In CSPM the strong color field inside the large cluster produces de-acceleration of the primary $q \bar q$ pair which can be seen as a thermal temperature by means of the Hawking-Unruh effect.    
The string percolation density parameter $\xi$ which characterizes the percolation clusters measures the initial temperature of the system. Since this cluster covers most of the interaction area, this temperature becomes a global temperature determined by the string density.
In this way at $\xi_{c}$ = 1.2 the connectivity percolation transition at $T(\xi_{c})$ models the thermal deconfinement transition.

We adopt the point of view that the experimentally determined universal chemical freeze-out temperature ($\it T_{f}$) is a good measure of the phase transition temperature, $T_{c}$ \cite{braunmun}. ${\langle p_{t}^{2}\rangle_{1}}$ is evaluated using Eq.(9) at $\xi_{c}$ = 1.2 with $\it T_{f}$ = 167.7 $\pm$ 2.6 MeV \cite{bec1}. This gives $ \sqrt {\langle {p_{t}^{2}} \rangle _{1}}$  =  207.2 $\pm$ 3.3 MeV which is close to  $\simeq$ 200 MeV used in a previous calculation of the percolation transition temperature \cite{pajares3}. This calibrates the CSPM temperature scale. 
The dynamics of massless particle production has been studied in QED2 quantum electrodynamics.
QED2 can be scaled from electrodynamics to quantum chromodynamics using the ratio of the coupling constants. Here the production time for a boson (gluon) is $\tau_{pro}= \frac {2.405\hbar}{mc^{2}} $ \cite{wong}. This gives $\tau_{pro} \sim ~$ 1.13 fm for central Au+Au collisions at $\sqrt{s_{NN}}=$200 GeV. 
The temperature obtained using Eq. (9) was $\sim$ 193.6 MeV for Au+Au collisions. 
 For Pb+Pb collisions the temperature is $\sim$ 262.2 MeV for 0-5$\%$ centrality, which is expected to be $\sim$ 35 $\%$ higher than the temperature from  Au+Au collisions \cite{eos}. A recent summary of the results from Pb+Pb collisions at the LHC has mentioned that the initial temperature increases at least by 30 $\%$ as compared to the top RHIC energy \cite{summ}.
Table I gives the CSPM  values  $\xi$, $\it T$, $\varepsilon$ and $\eta/s$ at $T/T_{c}$ =0.88, 1, 1.16 and 1.57. 

One way to verify the validity of extrapolation from RHIC to LHC energy is to compare the energy density expressed as $\varepsilon/T^4$ with the available lattice QCD results. Figure 3 shows a plot of $\varepsilon/T^4$  as a function of T/$T_{c}$. The lattice QCD results are from HotQCD Collaboration \cite{hotqcd}. It is observed that at LHC energy the CSPM results are in excellent agreement with the lattice QCD results. The lattice and CSPM results are available for T/$T_{c} < 2$. 
\begin{figure}[thbp]
\centering        
\vspace*{-0.2cm}
\resizebox{0.55\textwidth}{!}{
\includegraphics{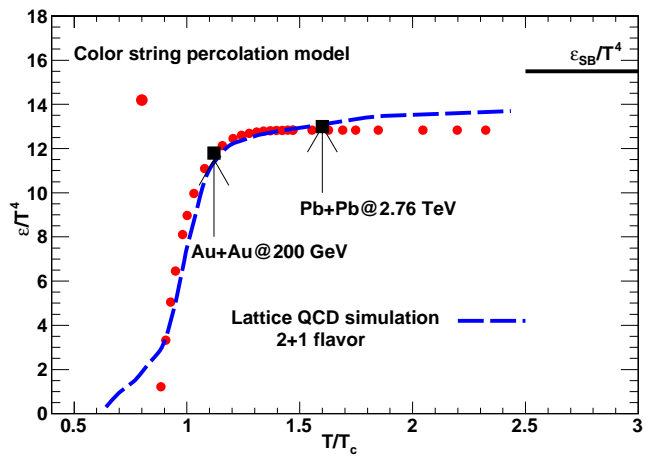}
}
\vspace*{-0.5cm}
\caption{ $\varepsilon/T^{4}$ as a function of T/$T_{c}$.The lattice QCD calculation is shown as dotted blue line \cite{hotqcd}. } 
\end{figure}     
\section{Shear Viscosity}
 The relativistic kinetic theory relation for the shear viscosity over entropy density ratio, $\eta/s$ is given by \cite{gul1,gul2}
\begin{equation}
\frac {\eta}{s} \simeq \frac {T \lambda_{mfp}}{5}     
\end{equation}
where T is the temperature and $\lambda_{mfp}$ is the mean free path given by
\begin{equation}
\lambda_{mfp} \sim \frac {1}{(n\sigma_{tr})}
\end{equation}
$\it n $ is the number density of an ideal gas of quarks and gluons and $\sigma_{tr}$ the transport cross section for these constituents. 

After the cluster is formed it behaves like a free gas of constituents. Eq. (10) can be applied to obtain the shear viscosity. In CSPM the number density is given by the effective number of sources per unit volume 
\begin{equation}
n = \frac {N_{sources}}{S_{N}L}
\end{equation}
 L is the longitudinal extension of the source, L = 1 $\it  fm $ \cite{pajares3}. The area occupied by the strings is related to $\xi$ through the relation $(1-e^{-\xi})S_{N}$. Thus the effective no. of sources is given by the total area occupied by the strings divided by the effective area of the string $S_{1}F(\xi) $. 
\begin{equation}
N_{sources} = \frac {(1-e^{-\xi}) S_{N}}{S_{1} F(\xi)} 
\end{equation}
 In general $N_{sources}$ is  smaller than the number of single strings. $N_{sources}$ equals the number of strings $N_{s}$ in the limit of $\xi $ = 0. 
The number density of sources from Eqs. (12) and (13) becomes
\begin{equation}
n = \frac {(1-e^{-\xi})}{S_{1}F(\xi) L}
\end{equation}
In CSPM the transport cross section $\sigma_{tr}$ is the transverse area of the effective string $S_{1}F(\xi)$. Thus $\sigma_{tr}$ is directly proportional to $F(\xi)$ and hence to $\frac {1}{T^{2}}$. The mean free path is given by
\begin{equation}
\lambda_{mfp} = {\frac {L}{(1-e^{-\xi})}} 
\end{equation}
For a large value of $\xi$ the $\lambda_{mfp}$ reaches a constant value.
$\eta/s$ is obtained from $\xi$ and the temperature
\begin{equation}
\frac {\eta}{s} ={\frac {TL}{5(1-e^{-\xi})}} 
\end{equation}
Well below $\xi_{c}$ , as the temperature increases, the string density increases and the area is filled rapidly  and $\lambda_{mfp}$ and $\eta/s$ decrease sharply. Above $\xi_{c}$, more than 2/3 of the area are already covered by strings, and therefore the area is not filling as fast and the relatively small decrease of $\lambda_{mfp}$ is compensated by the rising of temperature, resulting in a smooth increase of $\eta/s$. The behavior of $\eta/s$ is dominated by the fractional area covered by 
strings. This is not surprising because $\eta/s$ is the ability to transport momenta at large distances and that has to do with the density of voids in the matter.

\begin{figure}[thbp]
\centering        
\vspace*{-0.2cm}
\resizebox{0.55\textwidth}{!}{
\includegraphics{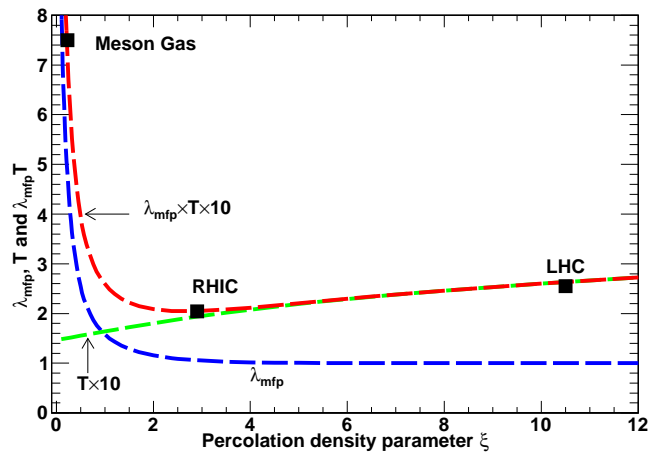}
}
\vspace*{-0.5cm}
\caption{$\lambda_{mfp}$ in fm ( blue line), temperature in GeV scaled by a factor of 10(green line) and $\lambda_{mfp} \times$ scaled T in GeV fm (red line) as a function of $\xi$. The minimum in $\eta/s$ is due to the combination of $\lambda_{mfp}$ and T. Meson gas \cite{meson}, RHIC and LHC points are also shown as solid black square.} 
\end{figure} 
\section{Results and Discussion}
Figure 4 shows a plot of $\lambda_{mfp}$, $\it T$ and $\lambda_{mfp}\times \it T$ as a function of $\xi$. Thus the product  T($\xi$)$\times \lambda_{mfp}$ will have a minimum in $\eta/s$. It has been shown that $\eta/s$ has a minimum at the critical point for various substances for example helium, nitrogen and water \cite{larry}. Thus the measurement of $\eta/s$ as a function of temperature can indicate the critical point in the QCD phase diagram with $\it T \sim $ 175-185 MeV. 

\begin{table}
\caption{The measured percolation density parameter $\xi$, temperature $\it$T, energy density $\varepsilon$  and $\eta/s$ for the meson gas \cite{meson};  the hadron to QGP transition;  Au+Au at 200 GeV  and Pb+Pb at 2.76 TeV (estimated). Au+Au is for 0-10$\%$ and Pb+Pb is for 0-5$\%$ central events.}
\vspace*{0.5cm}
\setlength{\tabcolsep}{2pt}
\begin{tabular}{|c|c|c|c|c|}\hline
System & $\xi$ & T (MeV) & $\varepsilon (GeV/fm^3)$ & $\eta/s$ \\ \hline
 Meson Gas & 0.22 & 150.0 &- & 0.76 \\ \hline
 Hadron to QGP & 1.2 & 167.7 $\pm 2.6$ & 0.94$\pm 0.07$& 0.240$\pm 0.012$ \\ \hline
Au+Au & 2.88$\pm0.09$ & 193.6 $\pm3.0$ & 2.27$\pm 0.16$ & 0.204$\pm 0.020$ \\ \hline
Pb+Pb & 10.56$\pm1.05$ & 262.2$\pm13.0$ & 8.32$\pm 0.83$ &0.260$\pm 0.026$ \\ \hline 
\end{tabular}
\end{table}
 
\begin{figure}[thbp]
\centering        
\vspace*{-0.2cm}
\resizebox{0.55\textwidth}{!}{
\includegraphics{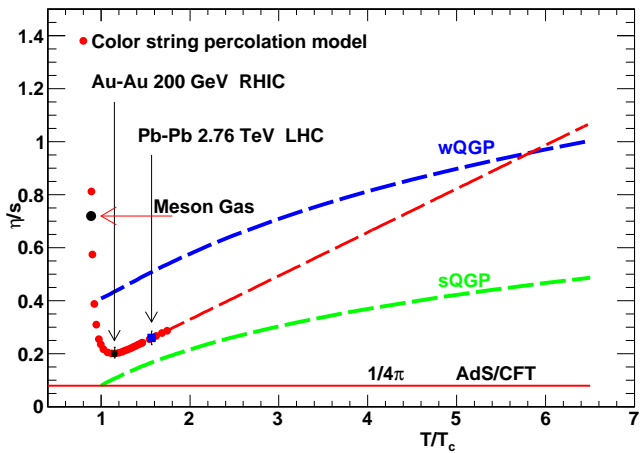}
}
\vspace*{-0.5cm}
\caption{$\eta/s$ as a function of T/$T_{c}$. Au+Au at 200 GeV for 0-10$\%$ centrality is shown as solid black square. wQGP and sQGP values are shown as dotted blue and green lines respectively \cite{gul1}. The estimated value for Pb+Pb at 2.76 TeV for 0-5$\%$ centrality  is shown as a solid blue square. The red dotted line represents the extrapolation to higher temperatures from the CSPM. The hadron gas value for $\eta/s$ $\sim$ 0.7 is shown as solid black circle at T/$T_{c} \sim $0.88 \cite{meson}.} 
\end{figure}       
 Figure 5 shows a plot of $\eta/s$ as a function of T/$T_{c}$. The estimated value of $\eta/s$ for Pb+Pb is also shown in Fig. 5 at T/$T_{c}$ = 1.57.
The lower bound shown in Fig. 5 is given by AdS/CFT \cite{kss}. These results from STAR and ALICE data show that the $\eta/s$ value is 2.5 and 3.3 times the KSS bound \cite{kss}. 

The theoretical estimates of $\eta/s$ has been obtained as a function of T/$T_{c}$ for both the weakly (wQGP) and strongly (sQGP) coupled QCD plasma are shown in Fig. 5 \cite{gul1}. It is seen that at the RHIC top energy  $\eta/s$ is close to the sQGP. Even at the LHC energy it follows the trend of the sQGP. By extrapolating the $\eta/s$ CSPM values to higher temperatures it is clear that $\eta/s$ could approach the weak coupling limit near $T/T_{c}$ $\sim$ 5.8. 
 The CSPM  $\eta/s$ value for the hadron gas is in agreement with the calculated value using measured elastic cross sections for a gas of pions and kaons \cite{prakash}. $\eta/s$ has also been obtained in several other calculations for pure glue matter \cite{toneev}, in the semi quark qluon plasma \cite{hidaka} and in quasiparticle description \cite{bluhm}. In pure SU(3)  gluodynamics a conservative upper bound for $\eta/s$ was obtained  $\eta/s$= 0.134(33) at $T=1.65 T_{c}$ \cite{meyer}.  In the quasiparticle approach also low $\eta/s \sim$ 0.2 is obtained for T $ > 1.05 T_{c}$ and rises very slowly with the increase in temperature \cite{peshier}. In CSPM also $\eta/s$ grows with temperature as 0.16T/$T_{c}$.

The CSPM model calculations have also successfully described  the elliptic flow and the nuclear modification factor at RHIC and LHC energies \cite{bautista}. In addition CSPM has determined the equation of state of the QGP and the bulk thermodynamic value of $\varepsilon/T^{4}$ and $s/T^{3}$ in excellent agreement with Lattice Gauge calculations \cite{eos}. This emphasizes the quantitative nature of the CSPM when applied to the data at $\sim$ 1 TeV scale. 
\section{Summary}
In summary the relativistic kinetic theory relation for shear viscosity to entropy density ratio  $\eta/s = \frac {1}{5}T$ $\lambda_{mfp}$ was evaluated as a function of the temperature using the measured transverse momentum spectra and the Color String Percolation Model.  The color suppression factor $F(\xi)$ was extracted from the transverse momentum spectrum of charged hadrons. We found $\eta/s$ = 0.204 $\pm 0.020$ at $T/T_{c}$ = 1.15 ( RHIC ) and $\eta/s$ =0.260 $\pm 0.020$ at $T/T_{c}$ = 1.57 (LHC). 
 In the phase transition region $\eta/s$ is 2-3 times the conjectured quantum limit for RHIC to LHC energies. 
The whole picture is consistent with the formation of a fluid with a low shear to viscosity ratio. The percolation framework provides us with a microscopic picture which predicts the early thermalization required for hydrodynamical calculations.

The minimum in $\eta/s$ can be studied as a function of the beam energy at RHIC that could locate the critical point/crossover in the QCD phase diagram seen in substances like helium, nitrogen and water \cite{teaney1,larry}. The accurate determination of $\eta/s$ is also important for the evaluation of another transport coefficient, the jet quenching parameter $\hat{q}$ \cite{muller,wang}.
\section{Acknowledgement}     
 This research was supported by the Office of Nuclear Physics within the U.S. Department of Energy  Office of Science under Grant No. DE-FG02-88ER40412.\\ J.D.D. thanks the support of the FCT/Portugal project PPCDT/FIS/575682004. C.P. was supported by the project FPA2011-022776 of MICINN the Spanish Consolider Ingenio 2010 program CPAN and Conselleria Education Xunta de Galicia.


\end{document}